# Detecting degeneracy and subtle broken-symmetry states of graphene at nanoscale

Si-Yu Li[1], Ya-Ning Ren[1], Yi-Wen Liu[1], Ming-Xing Chen[2], Hua Jiang[3], and Lin He[1,*]

[1]Center for Advanced Quantum Studies, Department of Physics, Beijing Normal University, Beijing, 100875, People's Republic of China

[2]College of Physics and Information Science, Hunan Normal University, Changsha, Hunan 410018, People's Republic of China

[3]College of Physics, Optoelectronics and Energy and Institute for Advanced Study, Soochow University, Suzhou, 215006, People's Republic of China

Correspondence and requests for materials should be addressed to L.H. (e-mail: helin@bnu.edu.cn).

**Measuring degeneracy and broken-symmetry states of a system at nanoscale requires extremely high energy and spatial resolution, which has so far eluded direct observation. Here, we realize measurement of the degeneracy and subtle broken-symmetry states of graphene at nanoscale for the first time. By using edge-free graphene quantum dots, we are able to measure valley splitting and valley-contrasting spin splitting of graphene at the single-electron level. Our experiments detect large valley splitting around atomic defects of graphene due to the coexistence of sublattice symmetry breaking and time reversal symmetry breaking. Large valley-contrasting spin splitting induced by enhanced spin-orbit coupling around the defects is also observed. These results reveal unexplored exotic electronic states in graphene at nanoscale induced by the atomic defects.**

Knowing degeneracy and exact nature of broken-symmetry states of a system is a central problem in physics. It also plays a vital role in understanding electronic properties of materials. Owing to the two-dimensional nature of graphene's structure and electrons, it is convenient to measure the degeneracy and exotic broken-symmetry states of graphene through a magneto-transport measurement of quantum Hall effect (QHE)[1-7]. As schematically shown in Fig. 1a and 1b, it is expected to observe quantum Hall conductance plateaus at values $4(n+1/2)e^2/h$ in pristine graphene ($n$ is integer, $e$ is electron, and $h$ is Planck's constant). Here 4 directly reflects the four-fold, including double-spin and double-valley, degeneracy of graphene. If the degeneracy of graphene is lifted and broken-symmetry states emerge, there should be QHE plateau not at the values $4(n+1/2)e^2/h$ (Fig. 1b shows one of the simplest case: the valley degeneracy is lifted in the zeroth Landau level)[4-7]. Therefore, the QHE provides a quite powerful method in detecting the degeneracy and broken-symmetry states of a system. However, the magneto-transport measurement lacks spatial resolution that limits its application at nanoscale. Atomic defects, such as single carbon vacancy[8] and adatoms[9], are almost unavoidable in graphene. An individual atomic defect could locally break the equivalence of two sublattices, which is expected to lift the degeneracies of graphene at nanoscale[10,11]. However, measuring the degeneracy and broken-symmetry electronic states around the atomic defects requires nanometer-scale spatial resolution, which is a key challenge to probe these interesting properties, leaving them unexplored in experiment up to now. Here, we realize the measurement of the degeneracy and subtle broken-symmetry states of graphene at the nanoscale. Giant valley splitting and valley-contrasting spin splitting are directly detected around the atomic defects of graphene at the single-electron level.

In our experiment, the measurement is realized by using a recently developed edge-free graphene quantum dots (QDs)[12,13], which can be generated by a combination of electrostatic and magnetic fields, as schematically shown in Fig. 1c and Fig. 1d. The magnetic fields quantize the continuous electronic spectrum of graphene into discrete Landau levels (LLs). The probing scanning tunneling microscopy (STM) tip, acting as a moveable top gate in the measurement[14-17], shifts quasiparticles in the region beneath the tip into the gaps between the LLs of the adjacent regions. This leads to edge-free confinement in graphene and generates almost equally-spaced orbital states in the QDs. Because the confinement is achieved without resorting to physical edges, both the two-fold valley and spin symmetries of the pristine graphene are preserved in the orbital states of the edge-free QDs[12,13], as shown in Fig. 1e. In the STM measurements, a single excess electron on the graphene QD needs to overcome the electrostatic energy $E_C = e^2/C$, where $C$ is capacitance of the QD and $e$ is the charge of an electron[18]. Therefore, it is expected to observe a series of quadruplets of charging peaks in scanning tunneling spectroscopy (STS) spectrum of pristine graphene, as schematically shown in Fig. 1e. Such a result has been demonstrated explicitly in previous studies[12,13] and also in our experiment, as shown in Fig. 2 and Figs. S1-S3. Each quadruplet belongs to a single orbital state of the edge-free QD and reflects the four-fold degeneracy of graphene[12,13,19,20]. When the degeneracy of graphene is partially lifted, for example, the valley degeneracy is lifted (Fig. 1f), then the quadruplet will split into two doublet. If the spin degeneracy in each valley is further lifted, then the four charging peaks of every quadruplet are no longer equally spaced (Fig. 1f). Therefore, it is convenient to know the degeneracy and measure the broken-symmetry states of graphene based on the four charging peaks of each quadruplet. By taking advantage of high spatial resolution of STM, our experiment, as shown subsequently, demonstrated that we can measure the degeneracy and broken-symmetry states of graphene at nanoscale.

The edge-free graphene QD can be moved by moving the STM tip, which enables us to measure electronic properties at any chosen position in graphene. Therefore, we carried out measurements around atomic defects to explore effects of the atomic defects on the four charging peaks of the quadruplet. Figure 2 shows our experimental results obtained in a graphene monolayer with atomic defects (including single carbon vacancy and adatoms, see Fig. S4 for details of sample preparation and Fig. S5 for Landau quantization measurement of the sample). According to the STM image (Fig. 2a), the studied area can be roughly divided into two regions, as separated by the dashed curve. Region 1 is free of defect and we can clearly observe atomic lattice of graphene in it. In region 2, the main feature in the STM image is triangular $\sqrt{3}\times\sqrt{3}R30°$ interference pattern, which is the characteristic feature of intervalley scattering induced by the atomic defects[8,9,21] (see Fig. S6 for fast Fourier transform images of the two regions). Besides the single carbon vacancy (marked by blue arrow in Fig. 2a), as revealed in the STM image, we identify several other atomic defects in the region 2 in our STS map (marked by black arrows in Fig. 2b). These defects exhibit tip-gate-induced ionization and show similar features as that reported previously for adatoms on graphene[22].

Therefore, they are attributed to adatoms trapped beneath the graphene sheet. The adatoms trapped beneath graphene are most possibly hydrogen atoms. In previous studies, the hydrogen adsorbing on graphene could induce a localized state in graphene[9], which was explicitly observed in our STS measurements (see Fig. S7). Moreover, the hydrogen atoms exist naturally in the sample synthesized by our method (they are generated in the dissociation process of both $CH_4$ and $H_2$ at high temperature) and they could intercalate graphene through defects in graphene[23]. The existence of both the single carbon vacancy and the hydrogen atoms results in large sublattice asymmetry of graphene (see Fig. S8) and, consequently, we observe clear triangular lattice in the STM image of the region 2.

Figure 2c and 2d show STS spectra recorded in different magnetic fields in the regions 1 and 2. Our STS spectra reveal well-defined Landau quantization of massless Dirac fermions[24-28], demonstrating explicitly that the topmost graphene sheet is electronically decoupling from the supporting substrate. Besides the well-defined LLs, we also observe quadruplets of charging peaks in the tunneling spectra, indicating the emergence of the edge-free graphene QD beneath the STM tip. According to our experiment, the first four charging peaks in the two regions exhibit quite different features. In the region 1, the four charging peaks are almost equally spaced. However, the quadruplets are obviously divided into two doublets in the spectra recorded in the region 2. The sublattice asymmetry in the region 2 breaks the inversion symmetry of graphene and lifts energy degeneracy of the *A* and *B* sublattices. In the quantum Hall regime, the broken symmetry of the graphene sublattices shifts the energies of the *n* =

0 LL in the *K* and *K′* valleys in opposite directions and, therefore, generates valley splitting in the $n = 0$ LL[29]. In our experiment, the first four charging peaks in Fig. 2c and 2d arise from the confinement of the $n = 0$ LL. When the valley degeneracy of graphene is lifted, every single quadruplet will be separated into two doublets, as schematically shown in Fig. 1f, and as demonstrated very recently in the edge-free graphene QDs on hexagonal boron nitride[13]. The above result indicates that the valley degeneracy in graphene is locally lifted in the region 2.

To clearly show how the atomic defects affect the energy spacing between the four charging peaks of a confined orbital state, we measured $\Delta E_{12}$, $\Delta E_{23}$, and $\Delta E_{34}$ (as defined in Fig. 1f) as a function of positions in the studied area shown in Fig. 2a. Figures 2e-2g show our experimental result measured at $B = 10$ T (see Fig. S9 for the voltage maps of the four charging peaks at 10 T). In the region 1, the $\Delta E_{12}$, $\Delta E_{23}$, and $\Delta E_{34}$ exhibit similar slight spatial variation and the difference between them is quite small. However, in the region 2, the $\Delta E_{23}$ shows a much strong spatial variation and becomes much larger than the $\Delta E_{12}$ and $\Delta E_{34}$. Such a result completely removes spatial variations of the $E_C$ as the origin of the observed features (the spatial variations of the $E_C$ will change the $\Delta E_{12}$, $\Delta E_{23}$, and $\Delta E_{34}$ simultaneously). The $\Delta E_{23}$ includes the charging energy $E_C$, the Zeeman-like splitting $E_Z$, and the valley splitting $E_v$, as schematically shown in Fig. 1f. The $E_C$ depends weakly on the recorded positions and will affect the $\Delta E_{12}$, $\Delta E_{23}$, and $\Delta E_{34}$ simultaneously. The $E_Z$ is expected to be independent of the measured positions ($E_Z = g\mu_B B \sim 1.1$ meV at $B = 10$ T with assuming the effective *g*-factor $g = 2$). Therefore, the obvious increase of the $\Delta E_{23}$ in the region 2 mainly reflects the valley splitting induced by the coexistence of the sublattice-symmetry breaking around the atomic defects and the magnetic fields.

To qualitatively explore the spatial variation of the valley splitting, we plot the valley splitting $E_v = \Delta E_{23} - E_{add}$ as a function of positions in Fig. 3a. Here $E_{add}$, including the $E_C$ and the $E_Z$, is the averaged energy spacing in the quadruplet obtained in the region 1 where the four charging peaks are almost equally spaced (see Fig. S10 for the measured $E_{add}$ as a function of magnetic fields). The maximum valley splitting at 10 T reaches about 43 meV. In literature, the valley splitting observed in the edge-free graphene QDs on hexagonal boron nitride is about 10 meV[13] and the largest valley splitting reported so far, which is induced by the coexistence of strain-induced pseudomagnetic field and external magnetic field, is only about 26 meV[30]. According to our experimental result, the observed valley splitting is closely related to the atomic defects in the studied area and can spatially extend about 3-5 nm around the defects.

Besides the predominantly valley splitting, we find that the energy spacings between two charging peaks of the two doublets, $\Delta E_{12}$ and $\Delta E_{34}$, are slightly different in the studied area (Fig. 2e and Fig. 2g). Such a result, which is unexpected and has never been reported previously in graphene, indicates that the spin splittings in the two valleys are not equal to each other, as schematically shown in Fig. 1f. To clearly show this phenomenon, we plot the spin splitting in the $K$ valley $E_S^K = \Delta E_{12} - E_{add}$ and in the $K'$ valley $E_S^{K'} = \Delta E_{34} - E_{add}$ as a function of positions in the studied area in Fig. 3b and 3c. Obviously, the two figures exhibit quite different features and there is an obvious valley-contrasting spin splitting: the $E_S^K$ is much larger than that in the $K'$ valley in the region 2. The maximum $E_s^K$ reaches about 8 meV at 10 T, whereas it only about 2.5 meV at 10 T in the *K'* valley. The above experimental results are further confirmed in other different magnetic fields. Figure 4a and the upper panel of Fig. 4b summarize the maximum valley splitting and the maximum spin splitting in the two valleys obtained around the atomic defects as a function of magnetic fields, respectively. In the lower panel of Fig. 4b, we show the difference of the maximum spin splitting in the two valleys $\Delta E_s$ obtained in our experiment, which is about 5 meV and depends weakly on the magnetic fields. All the experiments clearly demonstrate the giant valley splitting and valley-contrasting spin splitting around the atomic defects of graphene.

Our experimental observations can be understood within a theoretical framework that incorporates the effects of the sublattice-symmetry breaking and spin-orbit coupling (SOC) induced by the atomic defects in graphene[10,11,31-35]. In pristine graphene, the fourfold valley and spin degeneracy preserves and the large magnetic fields only slightly lift the fourfold degeneracy through Zeeman-like splitting, as shown in left panel of Fig. 4c. In such a case, the four charging peaks of every quadruplet are always equally spaced. The coexistence of the sublattice symmetry breaking and the large perpendicular magnetic fields will lift the valley degeneracy of the $n = 0$ LL in graphene[10,11,29]. This is especially the case around the atomic defects in our experiment. Therefore, every single quadruplet arising from the confinement of the $n = 0$ LL will be separated into two doublets (the middle panel of Fig. 4c), which is the predominant feature observed around the atomic defects (Fig. 2 and Fig. 3). By further taking into account the enhanced SOC introduced by the hydrogen adatoms, the spin degeneracy of graphene is further lifted and the spin splittings in the two valleys could be different[31-35], as schematically shown in the right panel of Fig. 4c.

To qualitatively understand the observed valley splitting and valley-contrasting spin splitting, we carried out tight-binding calculations by taking into account of the sublattice-symmetry breaking, the SOC and the magnetic fields (see supplemental materials for details of calculations). Figure 4d shows the obtained energies of the LLs. In the $n = 0$ LL, the fourfold valley and spin degeneracy is lifted by the coexistence of the sublattice symmetry breaking and the SOC. Importantly, the SOC generates different spin splittings in the two valleys. As a consequence, we observe valley splitting and valley-contrasting spin splitting around the atomic defects. According to our experimental result and theoretical analysis, the SOC strengths induced by the atomic defects in the $K$ and $K'$ valleys of the $n = 0$ LL are about 4 meV and 1.2 meV, respectively. The generated spin splittings at the Dirac points of the $K$ and $K'$ valleys are about 8 meV and 2.4 meV, respectively, which are about $10^4$ times larger than the expected SOC gap, ~ $8 \times 10^{-4}$ meV, at the Dirac point of pristine graphene[36,37]. Theoretically, hydrogen atoms hybridize directly with a carbon atom of graphene could lead to a significant enhancement of the SOC, which is predicted to be about 7 meV[38]. Previous transport measurement in weakly hydrogenated graphene observed an enhanced SOC strength of 2.5 meV[39]. Therefore, the detected SOC strength induced by the hydrogen atoms in this work is quite reasonable.

In summary, we realize the measurements of valley splitting and valley-contrasting spin splitting in graphene at the nanoscale and the single-electron level. Our experiment provides a general strategy to probe the degeneracy and nature of exotic broken-symmetry states at nanoscale in graphene. This strategy is quite robust and could be extended to other two-dimensional systems.

**Acknowledgements**

We thank Ka Shen, Haiwen Liu, Zhe Yuan and Ji Feng for helpful discussion. This work was supported by the National Natural Science Foundation of China (Grant Nos. 11674029, 11422430, 11374035), the National Basic Research Program of China (Grants Nos. 2014CB920903, 2013CBA01603). L.H. also acknowledges support from the National Program for Support of Top-notch Young Professionals, support from "the Fundamental Research Funds for the Central Universities", and support from "Chang Jiang Scholars Program".

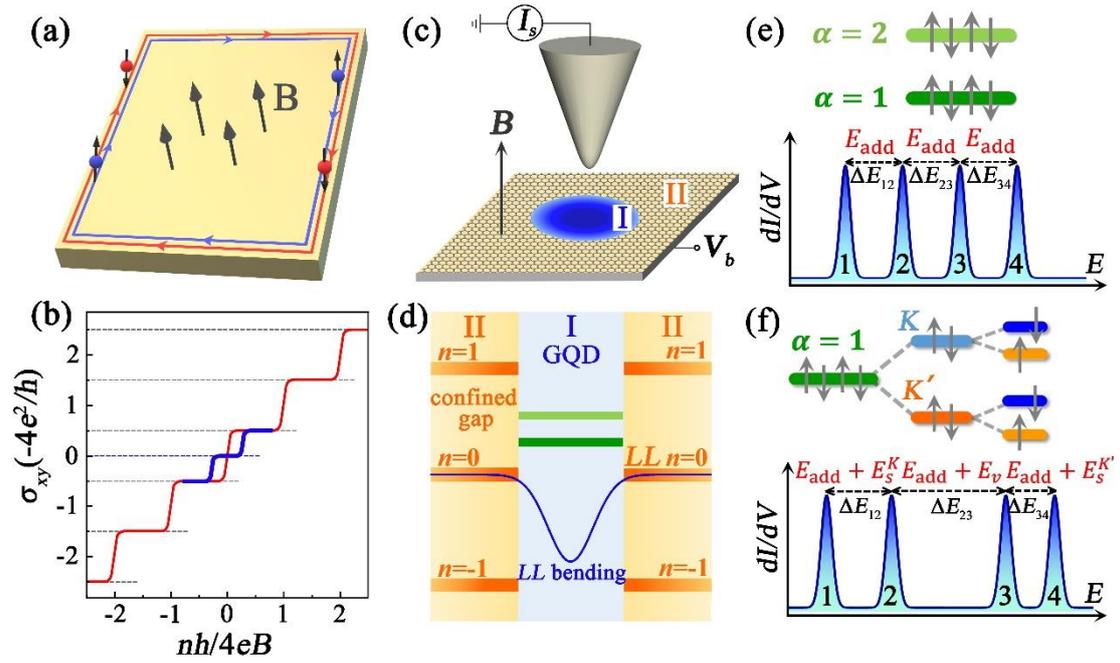

**Fig. 1. a.** Schematic diagram showing the QHE. **b.** The expected Hall conductivity in the integer quantum Hall states in graphene monolayer. The valley degeneracy is lifted in the zeroth Landau level in the blue curve. **c.** Sketch of the STM experiment in the presence of magnetic fields. The graphene monolayer is divided into two regions: one is the graphene region beneath the STM tip (region I), the other is the surrounding graphene region (region II). **d.** Schematic image of the tip-induced edge-free GQDs. Tip-induced electrostatic potential leads to LL bending in the region I. The gaps between the LLs in the region II provide the confined gaps, resulting in the edge-free GQDs (region I). **e.** Top: Schematic confined energy levels in the edge-free GQD with two orbital levels. Bottom: Schematic for a sequence of charging peaks in the $dI/dV$ spectra. **e.** Top: Schematic confined energy levels in the edge-free GQD. The orbital level $\alpha = 1$ exhibits the valley splitting $E_v$ and the spin splitting $E_s^k$, $E_s^{k'}$ in $K$ valley and $K'$ valley respectively. Bottom: Schematic for the corresponding charging peaks in $dI/dV$ spectrum.

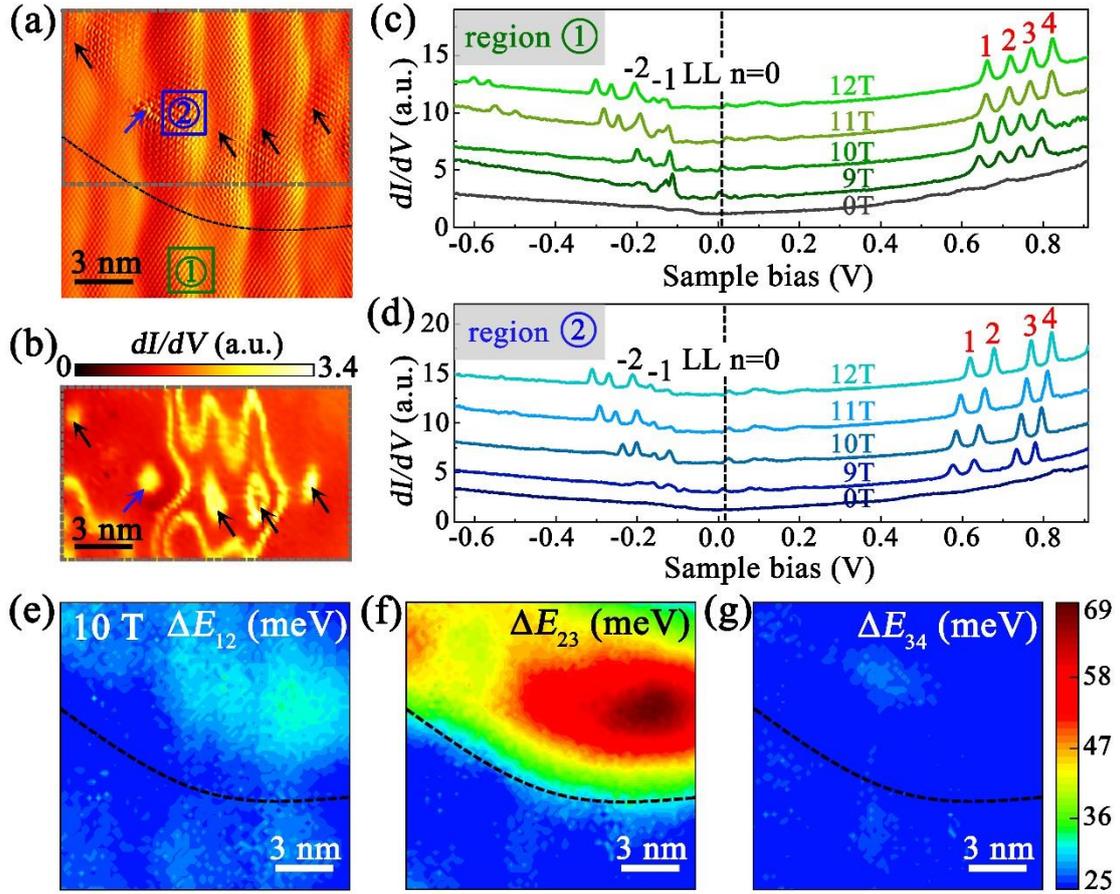

**Fig. 2. a.** A 15 nm × 15 nm STM image ($V_{sample}$ = 200 mV and $I$ = 0.4 nA) of graphene monolayer with atomic defects. The black dashed line roughly separates the lower regions without intervalley scattering (region 1) and the upper region with intervalley scattering (region 2). **b.** STS ($dI/dV$) map recorded in the region marked by grey dashed frame in panel A at the fixed sample bias 617 mV ($I$ = 0.4 nA). The arrows mark the positions of the atomic defects. **c, d.** STS spectra taken from the regions 1 and 2 in the graphene monolayer, respectively. The curves are offset on the Y axis for clarity and the LL indices are labeled by black numbers. Red numbers 1-4 mark the first four charging peaks. At negative voltage, some charging peaks are mixed with the LLs. **e-g.** 15 × 15 nm$^2$ energy maps for $\Delta E_{12}$, $\Delta E_{23}$, and $\Delta E_{34}$ with the magnetic field 10 T recorded in the same area shown in Fig. 2A. The black dashed line is the boundary separating the region 1 and 2, as defined in panel a.

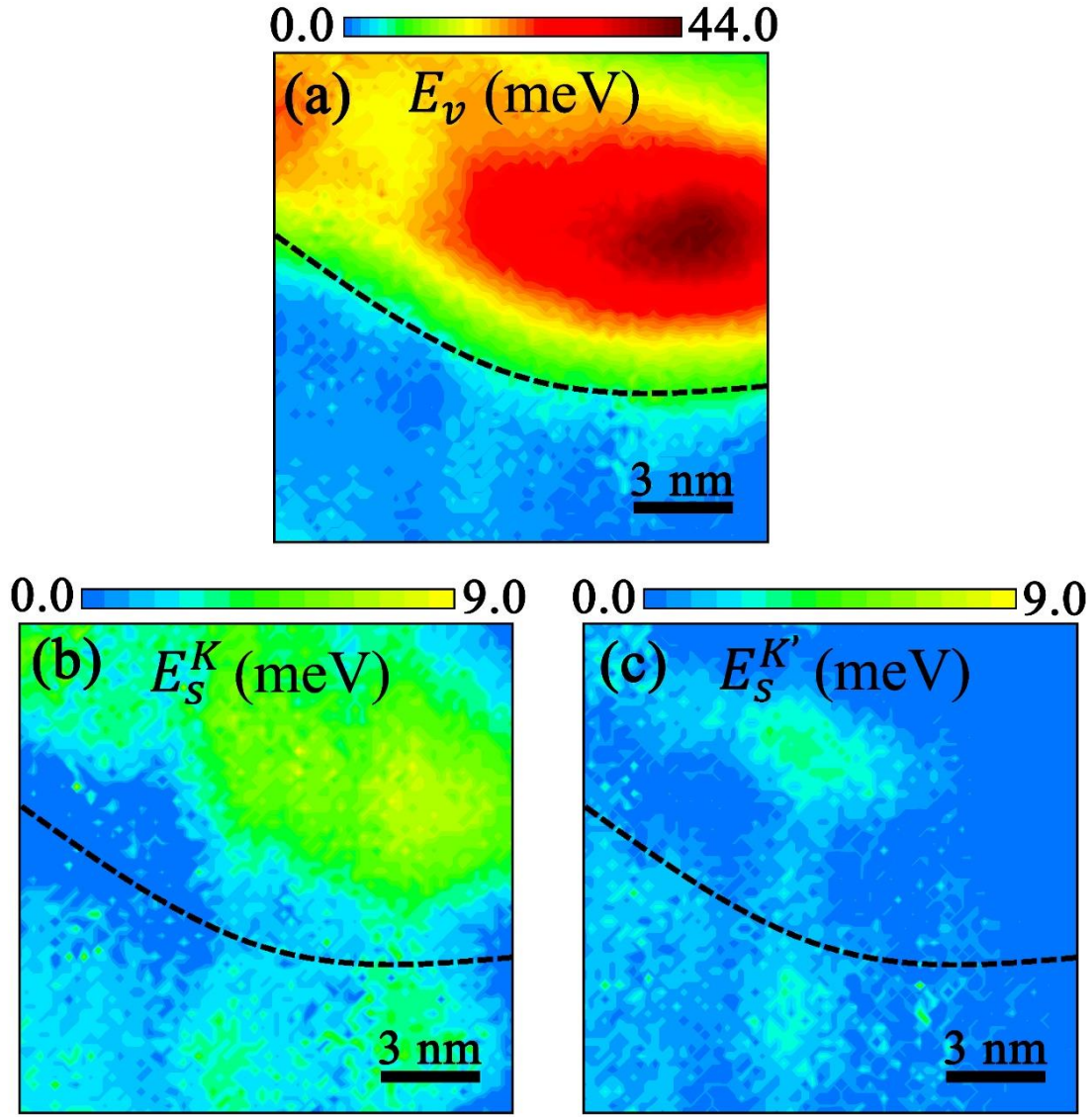

**Fig. 3. a-c.** 15 × 15 nm² energy maps for valley splitting $E_v$, spin splitting $E_s^k$ in the $K$ valley and spin splitting $E_s^{k'}$ in the $K'$ valley in the same area as Fig. 2a with the magnetic field 10 T. The black dashed line is the boundary separating the region 1 and 2.

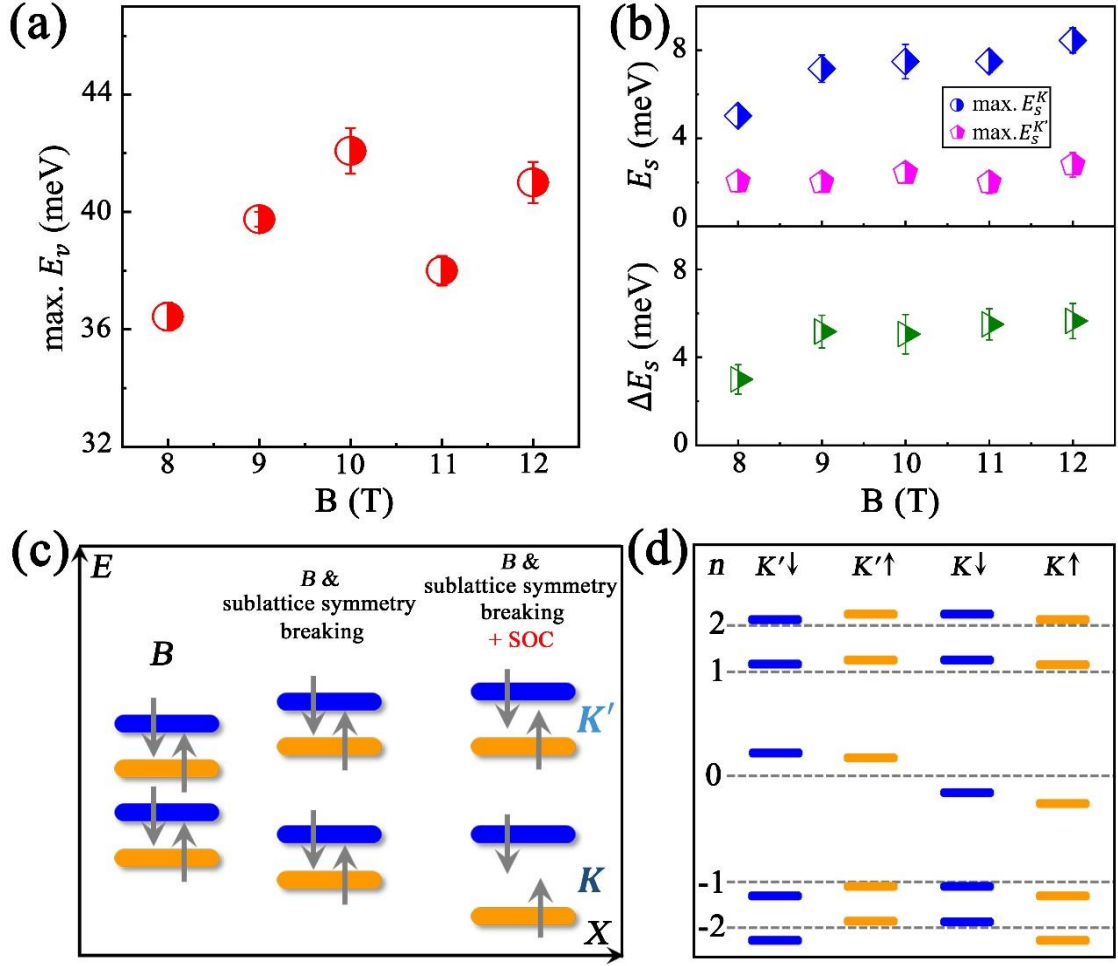

**Fig. 4. a.** The maximum valley splitting $E_v$ obtained around the atomic defects with different magnetic fields. **b.** Upper panel: The maximum $K$-valley spin splitting $E_s^k$ and maximum $K'$-valley spin splitting $E_s^{k'}$ obtained around the atomic defects with different magnetic fields. Lower panel: the difference $\Delta E_s$ between the maximum $E_s^k$ and the maximum $E_s^{k'}$ obtained around the atomic defects as a function of magnetic fields. **c.** Left: in pristine graphene, the fourfold degeneracy is almost not lifted (the large magnetic fields only slightly lift them through Zeeman-like splitting). Middle: the valley degeneracy in $n = 0$ LL is lifted due to the sublattice-symmetry breaking. Right: the fourfold valley and spin degeneracy in the $n = 0$ LL is lifted because the coexistence of the sublattice symmetry breaking and the SOC. Moreover, the SOC generates valley-

contrasting spin splitting in the two valleys of graphene. **d.** Schematic image for the electron energies in different Landau levels considering the sublattice symmetry breaking and SOC in graphene. The dashed lines mark the Landau levels in pristine graphene monolayer.